\RequirePackage[T1]{fontenc}
\documentclass[12pt]{article}
\pdfoutput=1 % if your are submitting a pdflatex (i.e. if you have images in pdf, png or jpg format)
\usepackage[height=8.85in,width=6.45in]{geometry}

\usepackage[utf8]{inputenc}
\usepackage{amsmath}
\usepackage{amssymb}
\usepackage{mathtools}
\numberwithin{equation}{section}
\usepackage{slashed}
\usepackage{braket}
\usepackage[svgnames]{xcolor}
\usepackage[colorlinks,linktocpage=true,citecolor=DarkGreen,linkcolor=FireBrick]{hyperref}
\usepackage{cite}
\usepackage{graphicx}
\usepackage{tikz}
\usepackage{tikz-cd}
\usepackage{times}
\usepackage[scaled]{couriers}
\usepackage{bm}
\usepackage{subfig}
\usepackage{mathrsfs}

\usepackage{xcolor}
\usepackage{mdframed}

\renewenvironment{figure}[1][]{
  \begin{originalfigure}[#1]
    \begin{mdframed}[linecolor=black!0,backgroundcolor=black!1]
}{
    \end{mdframed}
  \end{originalfigure}
}
\DeclareMathOperator{\Tr}{Tr}
\DeclareMathOperator{\tr}{tr}

\def\index{\mathop{\mathrm{index}}}

\def\sign{\mathop{\mathrm{sign}}}
\usepackage{slashed}

\def\vev#1{\langle #1 \rangle}

\def\cD{{\cal D}}

\def\cL{{\cal L}}

\def\cO{{\cal O}}

\def\bR{{\mathbb R}}

\def\bZ{{\mathbb Z}}

\def\U{\mathrm{U}}

\def\beq#1\eeq{\begin{align}#1\end{align}}

\def\blue#1{{\color{blue}#1}}

\def\o{\overline } 
\def\w{\widetilde } 

\def\Ch{\overline{\gamma}}
\def\d{{\rm d}}
\def\i{{\mathsf i}}

\def\ch{{\rm ch}}

\def\trg{G}

\begin{document}

\begin{titlepage}

\begin{flushright}
TU-1119
\end{flushright}

\vskip 3cm

\begin{center}

{\Large \bfseries Atiyah-Patodi-Singer index theorem from axial anomaly}

\vskip 1cm
Shun K. Kobayashi and Kazuya Yonekura
\vskip 1cm

\begin{tabular}{ll}
Department of Physics, Tohoku University, Sendai 980-8578, Japan
\end{tabular}

\vskip 1cm

\end{center}

\noindent
We give a very simple derivation of the Atiyah-Patodi-Singer (APS) index theorem and its small generalization by using the path integral of
massless Dirac fermions. It is based on the Fujikawa's argument for the relation between the axial anomaly and the
Atiyah-Singer index theorem, and only a minor modification of that argument is sufficient to show the APS index theorem.
The key ingredient is the identification of the APS boundary condition and its generalization as physical state vectors in the Hilbert space
of the massless fermion theory. The APS $\eta$-invariant appears as the axial charge of the physical states.

\end{titlepage}

\setcounter{tocdepth}{2}
%\tableofcontents

\newpage

\tableofcontents

\section{Introduction}
The Atiyah-Patodi-Singer (APS) index theorem~\cite{Atiyah:1975jf} is a generalization of the Atiyah-Singer (AS) index theorem
to the case of manifolds with boundaries. Let us first explain the theorem.

\begin{figure}
\centering
\includegraphics[width=0.5\textwidth]{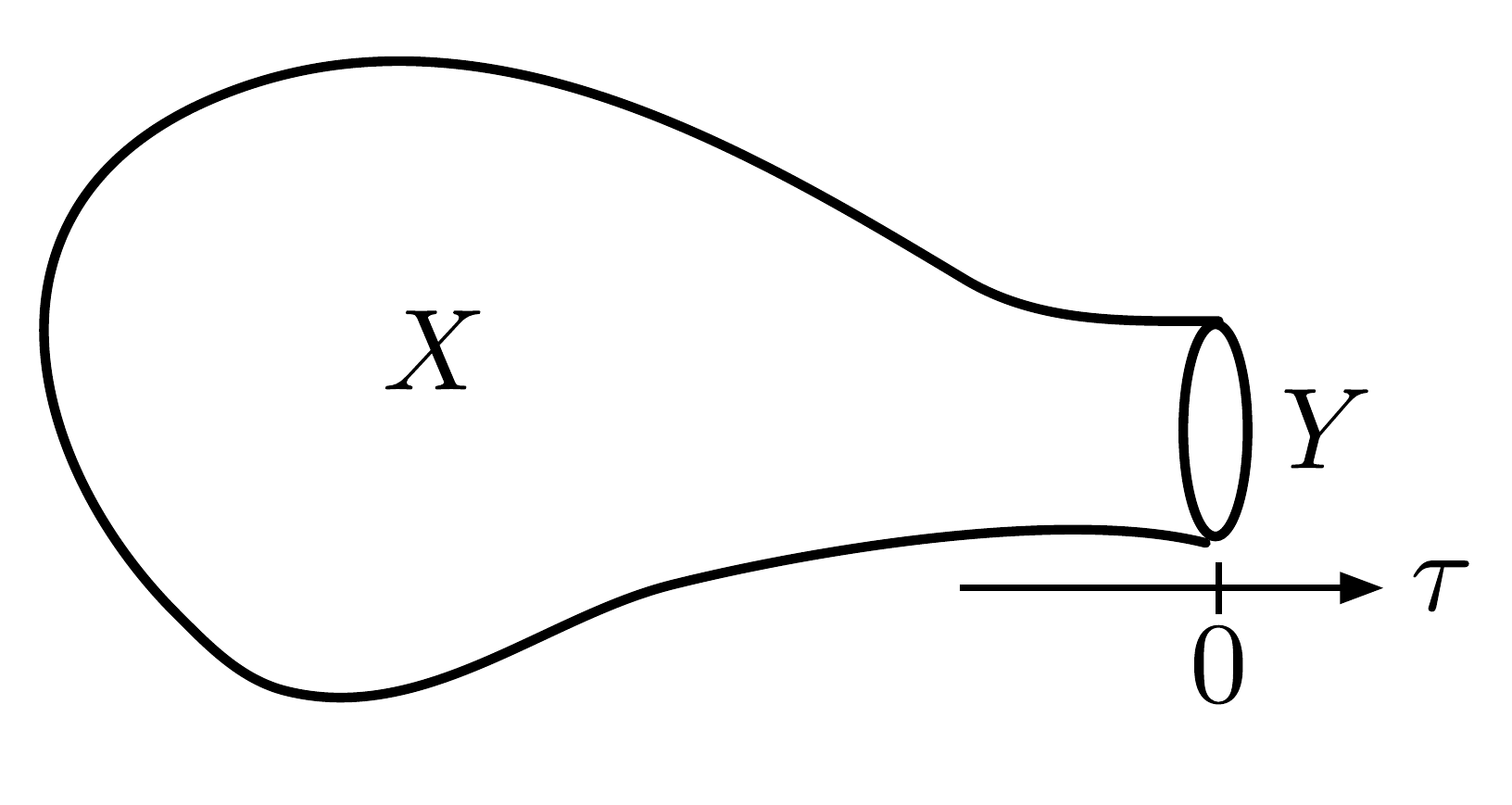}
\caption{A manifold $X$ with boundary $Y$. In the region near the boundary, 
we have a coordinate $\tau$ whose direction is orthogonal to the boundary.  \label{fig:simple}}
\end{figure}

We consider a manifold $X$ with boundary $\partial X = Y$, and we assume that the region near the boundary is given by $(-\epsilon, 0] \times Y$
where $\epsilon>0$ is some positive constant and $0 \in (-\epsilon,0] $ corresponds to the boundary of $X$. 
See Figure~\ref{fig:simple} for the situation. 
We assume that the background metric is also of the product form $(-\epsilon,0] \times Y$, and the gauge field
is independent of $\tau \in (-\epsilon,0]$ and has no $\tau$-component near the boundary.
Then we consider a Dirac operator $\cD_X = \i \gamma^\mu D_\mu$ on $X$ (where $\i =\sqrt{-1}$). 
We assume that there is a chirality operator $\Ch$ (which is usually denoted as $\gamma_5$
in four dimensions) with the usual properties that $\Ch$ is hermitian, $\Ch^2=1$ and $\{\Ch, \cD_X \}:=\Ch\cD_X + \cD_X \Ch=0$. 
We also impose some appropriate boundary condition, called the APS boundary
condition, whose physical interpretation will be discussed later in this paper. Under the APS boundary condition,
the Dirac operator $\cD_X$ is self-adjoint and has a well-defined spectrum. 
Then we can define the index $\index \cD_X$ in the usual way as in the AS index theorem
as the difference of the numbers of zero modes of $\cD_X$ with positive chirality $\Ch=+1$ and negative chirality $\Ch=-1$.

The APS index theorem involves 
the APS $\eta$-invariant defined as follows. Let us take a coordinate $\tau \in (-\epsilon,0]$ near the boundary.
Then, the Dirac operator can be written near the boundary as
\beq
\cD_X = \i \gamma^\tau \left( \frac{\partial}{\partial \tau} +  \widetilde \cD_Y \right),
\eeq
where $\gamma^\tau$ is the gamma matrix in the direction $\tau$, and 
$\widetilde \cD_Y =  \gamma^\tau \gamma^i D_i$ where the index $i$ is summed over the coordinates of $Y$.
We notice that $ \widetilde \cD_Y$ commutes with $\Ch$, so we can restrict $\widetilde \cD_Y$ to the subspace
with positive chirality $\Ch=+1$. We denote that operator as $\cD_Y$,
\beq
\cD_Y:= \widetilde \cD_Y|_{\text{subspace }\Ch=+1}.
\eeq
More explicitly, in some basis, we have 
\beq
\Ch=\begin{pmatrix} I & 0 \\ 0 & - I \end{pmatrix}, \qquad \gamma^\tau = \begin{pmatrix} 0 & I \\ I & 0 \end{pmatrix},\qquad 
\widetilde \cD_Y = \begin{pmatrix} \cD_Y & 0 \\ 0 & - \cD_Y \end{pmatrix},
\eeq
where $I$ is the identity matrix of the appropriate size, and in the last expression for $\widetilde \cD_Y$ we have used the fact that $\widetilde \cD_Y$ anticommutes with $\gamma^\tau$
and commutes with $\Ch$. The APS $\eta$-invariant of the operator $\cD_Y$ is defined in terms of the eigenvalues $\lambda_i$
of $\cD_Y$ as
\beq
\eta(\cD_Y) = \frac12 \left( \sum_i \sign(\lambda_i) \right)_{\rm reg}, \label{eq:intro_eta}
\eeq
where the sum is over all eigenmodes of $\cD_Y$, and 
$\sign(\lambda)$ is defined to be $\sign(\lambda)=\lambda/|\lambda|$ if $\lambda \neq 0$.
For the case $\lambda=0$, the value of $\sign(\lambda)$ depends on the precise definition of the APS boundary condition.
For example, in a certain choice of the APS boundary condition, we set $\sign(0)=+1$.
The subscript ${\rm reg}$ means that we need some regularization to make the infinite sum well-defined.

Now we can state the APS index theorem.
The index of the Dirac operator $\cD_X$ on $X$ with the APS boundary condition is given by
\beq
\index{\cD_X} = \int_X G + \eta(\cD_Y),
\eeq
where $G$ is the index density, which means that if the boundary is empty $Y = \varnothing$,
the usual AS index is given by the integral of $G$ over $X$.
The effect of the presence of the boundary is that we have the boundary contribution $\eta(\cD_Y)$ in the formula. 

From the APS formula, we can see that the APS $\eta$-invariant is roughly (or perturbatively) equal to the Chern-Simons invariant.
The reason is as follows. Since $\index{\cD_X}$ is an integer, let us neglect it for the purpose of seeing the continuous dependence of
$\eta(\cD_Y)$ on the background metric and gauge field. Also, let us suppose for simplicity that $G$ is a total derivative as
$G = \d  I_{d-1}$. Here $ I_{d-1}$ is a $(d-1)$-form (where $d=\dim X$) which we may call Chern-Simons form.
By using the Stokes theorem, we get $\eta(\cD_Y) \sim -  \int_X G \sim - \int_Y  I_{d-1}$.
The integral $\int_Y I_{d-1}$ is what is usually called the Chern-Simons invariant, at least if we neglect topological issues.
In fact, $\eta(\cD_Y)$ can be regarded as a nonperturbative version of the Chern-Simons invariant~\cite{AlvarezGaume:1984nf}
and appeared in global anomaly formulas~\cite{Witten:1985xe} which describe some nonperturbative anomalies.
Perturbative anomaly inflow~\cite{Callan:1984sa} in terms of Chern-Simons invariants 
can be generalized to nonperturbative anomalies~\cite{Witten:1999eg,Witten:2015aba,Witten:2019bou}, 
and it is possible to give a systematic description of anomalies
in terms of nonperturbative anomaly inflow and the $\eta$-invariant; see \cite{Witten:2019bou} for a systematic discussion. 
Therefore, the APS $\eta$-invariant has various applications to anomalies. For a sampling of recent applications,
see e.g. \cite{Witten:2016cio,Tachikawa:2018njr,Hsieh:2018ifc,Garcia-Etxebarria:2018ajm,Wang:2018qoy,Hsieh:2019iba,
Davighi:2019rcd,Wan:2019fxh,Hsieh:2020jpj,Davighi:2020uab}.

The APS index theorem is one of the most important theorems related to the APS $\eta$-invariant,
so it is desirable to understand this theorem as much as possible. 
The original proof is mathematically rigorous, but it is technically complicated. 
See \cite{Fukaya:2017tsq,Dabholkar:2019nnc,Fukaya:2019qlf,Ivanov:2020fsz,Fukaya:2020tjk}
for work which studies this theorem further from different points of view. 

As discussed by Fujikawa~\cite{Fujikawa:1979ay,Fujikawa:1980eg}, 
the AS index theorem can be understood by the axial anomaly of the path integral measure in massless fermion theories. 
We point out that Fujikawa's result can be used easily to give a physical derivation of the APS index theorem.
Our derivation is based on an axial symmetry transformation $\alpha(x)$ which depends on the
position of spacetime, $x \in X$. We will see that our method can bypass technically hard computations,
after noticing the physical interpretation of the APS boundary condition and the $\eta$-invariant.

\section{Axial $\U(1)$ charges and physical states}\label{sec:eta}
The APS $\eta$-invariant is the most important quantity in the APS index theorem.
Thus we review one of the several contexts in which the $\eta$-invariant appears naturally in physics.
It appears as the axial $\U(1)$ charge of the vacuum of massless fermions. (See e.g.~\cite{Lott:1984tn}.)

We consider a theory of massless Dirac fermions whose Lagrangian in the Lorentz signature metric is given by
\beq
\mathcal{L}_{\text{Lor}} = - \overline{\Psi} \gamma^\mu D_\mu\Psi,
\eeq
where $\gamma^\mu$ are gamma matrices, $D_\mu$ are covariant derivatives, $\Psi$ is the fermion field,
and $\overline{\Psi} = \Psi^\dagger (\i \gamma^0)$. The subscript ${\rm Lor}$ is put to emphasize that we are working in the Lorentz signature metric.
In addition, we assume that there exists a chirality operator $\Ch$ with the properties 
\beq
\Ch^2=1, \quad \{\Ch, \gamma^\mu\}=0, \quad \Ch^\dagger= \Ch.
\eeq
Therefore, there is an axial symmetry $\Psi \to e^{\i \alpha \Ch} \Psi$ at the classical level.

\subsection{The axial $\U(1)$ charge of the vacuum as the $\eta$-invariant }\label{subsec:detailq}

Let us consider the theory on a manifold $\bR \times Y$, where $Y$ is space and $\bR$ is time.
We take the time coordinate $t \in \bR$.
The Lagrangian can be rewritten as 
\beq
\mathcal{L}_{\text{Lor}}  = \i \Psi^\dagger \partial_t \Psi - \Psi^\dagger \w \cD_Y \Psi,
\eeq
where $\w \cD_Y = (\i \gamma^0) \gamma^i D_i$ in which the index $i$ is summed over the coordinates of the space $Y$.
We assume that all the background fields (i.e. the metric and gauge field) are independent of the time $t$,
and the gauge field does not have a component in the time direction.
The Hamiltonian density is 
$
    \mathcal{H}_Y = \Psi^\dagger \w {\mathcal D}_Y \Psi.
$

Because $\w {\mathcal D}_Y$ and $\overline{\gamma}$ commute, there are simultaneous eigenstates of them.
First we take eigenstates with $\Ch=+1$:
\beq
	\w {\mathcal D}_Y \Psi_{+,i} &= \lambda_{i}\Psi_{+,i},\\
	\overline{\gamma} \Psi_{+,i} &=+ \Psi_{+,i}.
\eeq
For convenience, and also to make the notation consistent with later sections, let us define $\gamma^\tau := \i \gamma^0$.
Then we consider $\gamma^\tau \Psi_{+,i}$. Because $\{\w{\mathcal D}_Y, \gamma^\tau\}=0$ and $\{\overline{\gamma}, \gamma^\tau\}=0$, we get
\beq
	\w{\mathcal D}_Y\left[\gamma^\tau \Psi_{+,i}\right]
	    & = -\gamma^\tau \w{ \mathcal D}_Y \Psi_{+,i} = - \lambda_{i}\left[\gamma^\tau \Psi_{+,i}\right],\\
	\overline{\gamma}\left[\gamma^\tau \Psi_{+,i}\right]
	    &=-\gamma^\tau\overline{\gamma} \Psi_{+,i}=  - \left[\gamma^\tau \Psi_{+,i}\right].
\eeq
Thus we can expand $\Psi$ by $\Psi_{+,i}$ and $\Psi_{-,i} = \gamma^\tau \Psi_{+,i}$ as
\beq
	\Psi = \sum_i \left[A_{+,i}\Psi_{+,i}+A_{-,i}\Psi_{-,i}\right]. \label{eq:PsiExpand}
\eeq
The canonical anticommutation relations are
\beq
\{A_{+,i}, A_{+,j}^\dagger\} = \delta_{ij}, \qquad \{A_{-,i}, A_{-,j}^\dagger\} = \delta_{ij}, \qquad \text{Others}=0.
\eeq
The Hamiltonian is\footnote{Throughout the paper, we omit to write the volume form in integrals over space (or spacetime). 
For example, $\int_Y = \int_Y \sqrt{g_Y} \d^d y$ for the metric $g_Y$ on $Y$.}
\beq
	H_Y 
		&= \int_{Y} \Psi^\dagger \w{\mathcal D}_Y \Psi  +\text{(const.)}
				=\sum_{i}\lambda_i\left[ N_{+,i} - N_{-,i}\right] +\text{(const.)}.
\eeq
where $N_{\pm,i} := A^\dagger_{\pm,i}A_{\pm,i}$. The axial charge is
\beq
   Q_A  
        %&
        = \int_{Y}  \Psi^\dagger \overline{\gamma}\Psi  +\text{(const.)}   %\notag\\
        %&
        =\sum_{i}\left[\left(N_{+,i}-\frac{1}{2}\right) - \left(N_{-,i}-\frac{1}{2}\right)\right]. \label{eq:Acharge} 
\eeq
In the last expression, we have chosen the constant to be such that the quantization
of each mode gives states with charge $\pm \frac12$. This definition turns out to be the correct one.\footnote{
One way to see that this is the correct definition is as follows. If we consider the vector $\U(1)$ symmetry
rather than the axial $\U(1)$ symmetry, we need the same prescription
as above so that the vacuum has charge zero under the vector $\U(1)$ symmetry.
Since the vector $\U(1)$ symmetry is anomaly-free, we expect that the correct prescription should give charge zero to the vacuum. }

\paragraph{The axial $\U(1)$ charge of the vacuum.}
For simplicity, we assume that there is no zero modes of the operator $\w \cD_Y$ until Sec.~\ref{sec:general}
where the most general case will be discussed.

We separate positive and negative eigenvalues, because we want to find the vacuum. The Hamiltonian is
\beq
	 H_Y = \sum_i{}^>\lambda_{i}^>\left[N_{+,i}^>-N_{-,i}^>\right] +\sum_i{}^<\lambda_{i}^<\left[N_{+,i}^<-N_{-,i}^<\right] +\text{(const.)},
\eeq
where
$\lambda^>_{i} >0 $ and $ \lambda^<_{i} < 0$, and
the summation in $\sum_i{}^{>}$ or $\sum_i{}^{<}$ runs over eigenstates which have positive or negative eigenvalues, respectively.
The vacuum $\ket{\Omega}$ has the minimum energy, so it satisfies the following equations:
\beq
	N_{+,i}^{>} \ket{\Omega}= N_{-,i}^{<} \ket{\Omega} = 0
	 ,\qquad N_{+,i}^{<} \ket{\Omega} = N_{-,i}^{>} \ket{\Omega}=1.
\eeq
These conditions characterize the vacuum.

The axial $\U(1)$ charge can be written as
\beq
	Q_A &= \sum_{i}{}^>\left[\left(N_{+,i}^>-\frac{1}{2}\right) - \left(N_{-,i}^>-\frac{1}{2}\right)\right]
		+ \sum_{i}{}^<\left[\left(N_{+,i}^<-\frac{1}{2}\right) - \left(N_{-,i}^<-\frac{1}{2}\right)\right].
\eeq
Therefore the axial $\U(1)$ charge of the vacuum is
\beq
    Q_A\ket{\Omega} 
		&=\left\{\sum_{i}{}^>\left[-\frac{1}{2}-\frac{1}{2} \right] + \sum_{i}{}^<\left[\frac{1}{2}+\frac{1}{2}\right]\right\}\ket{\Omega} \notag\\
		&= - \left\{\sum_{i}{}^> 1 - \sum_{i}{}^<   1  \right\}\ket{\Omega}\notag\\
		&= -2\eta(\cD_Y) \ket{\Omega}, 
\eeq
where $\eta(\cD_Y)$ is the APS $\eta$-invariant
\beq
	\eta(\cD_Y) = \frac{1}{2}\sum_i \text{sign}(\lambda_i)
		=\frac{1}{2}\left\{\sum_i{}^>  1   - \sum_i{}^<  1    \right\}. \label{eq:APSeta}
\eeq
This is the same definition as given in \eqref{eq:intro_eta}.
Therefore we get
\beq
   Q_A \ket{\Omega} =-2\eta(\cD_Y) \ket{\Omega}.\label{eq:Q=-2eta}
\eeq

\subsection{Physical states and boundary conditions}
Let us recall some facts about path integrals. They are very elementary, but are crucial for the purpose of the present paper.
To make the relation with the later sections clear, we perform Wick-rotation of the time coordinate as $\tau = \i t$.
Thus we consider the space $\bR \times Y$ with the Euclidean signature metric.

If we consider a time evolution from one state $\ket{\alpha}$ at the Euclidean time $\tau = -T$ to another state $\ket{\beta}$ at $\tau =0$,
it is described by the path integral as
\beq
\bra{\beta} e^{- T H} \ket{\alpha} = \int [\cD \Psi][ \cD\o \Psi]\exp\left( - \int_{[-T,0] \times Y} \cL_{\rm Euc} \right), \label{eq:T_evolve}
\eeq
where $H=H_Y$ is the Hamiltonian, and $ \cL_{\rm Euc}$ is the Lagrangian in the Euclidean signature metric, which is obtained from $\cL_{\rm Lor}$
as $\cL_{\rm Euc} = - \cL_{\rm Lor}$ after setting $t=-\i \tau$.
The information of the states $\bra{\beta}$ and $\ket{\alpha}$ are incorporated as the boundary conditions at $\tau=-T$
and $\tau = 0$, respectively. For example, this fact can be seen more easily in the ordinary quantum mechanics
described by $( x,  p)$. In that case, if we consider $\bra{x_1} e^{-TH} \ket{x_0}$,
then we perform the path integral with the boundary condition that $x(\tau=-T) = x_0$ and $x(\tau=0) = x_1$.

For fermions, the relation between a physical state
$\bra{\beta}$ and the corresponding boundary condition $\beta$ at $\tau=0$ can be worked out 
as in \cite{Yonekura:2016wuc}. Roughly it can be described as follows.
Let $L_\xi (\Psi)$ be a linear functional of $\Psi$ at the boundary
given by $L_\xi (\Psi)=\int_Y \xi^\dagger \Psi $. We can take $\xi$ to be one of the eigenmodes $\Psi_{\pm,i}$ of $\w \cD_Y$.
Then we have the correspondence
\beq
\begin{array}{llll}
\bra{\beta}L_\xi(\Psi)=0 & \Longrightarrow & L_\xi(\Psi)=0 & \text{(Dirichlet for the mode $\xi$)}, \\
\bra{\beta}L_\xi(\Psi)^\dagger =0 & \Longrightarrow & L_\xi(\Psi)=\text{free} & \text{(Neumann for the mode $\xi$)}.  
\end{array} \label{eq:BCandState}
\eeq
This is the basic dictionary between physical states and boundary conditions.

The path integral expression \eqref{eq:T_evolve} has a natural generalization when we are working in the Euclidean signature metric.
Let us consider a manifold $X$ which has a boundary $Y$ as in Figure~\ref{fig:simple}.
As we mentioned above, boundary conditions correspond to physical states.
So let us consider a boundary condition corresponding to a state $\bra{\beta}$.
The path integral without taking into account the boundary condition gives a state vector in the Hilbert space on $Y$. 
We denote this state vector as $\ket{X}$.
Then the path integral with the boundary condition $\bra{\beta}$ is actually computing the quantity
\beq
\bra{\beta} X\rangle = \int [\cD\Psi][ \cD\o \Psi]\exp\left( - \int_{X} \cL_{\rm Euc} \right).
\eeq
In the case of $X'= [-T,0] \times Y$, the state vector $\ket{X'}$ corresponds to the part $e^{- T H} \ket{\alpha}$. 
But for $X$ with $\partial X = Y$, there is no initial state.\footnote{It is analogous to the Hartle-Hawking no-boundary boundary condition,
although we treat the metric just as a background field.}

For the purpose of this paper, the important fact shown in \cite{Yonekura:2016wuc} is that the APS boundary condition
corresponds to the vacuum state $\bra{\Omega}$ of massless fermions. We will use this fact in our derivation of the APS index theorem.
The mathematical definition of the APS boundary condition is given by using \eqref{eq:BCandState} when $\bra{\beta} = \bra{\Omega}$.

\begin{figure}
\centering
\includegraphics[width=0.5\textwidth]{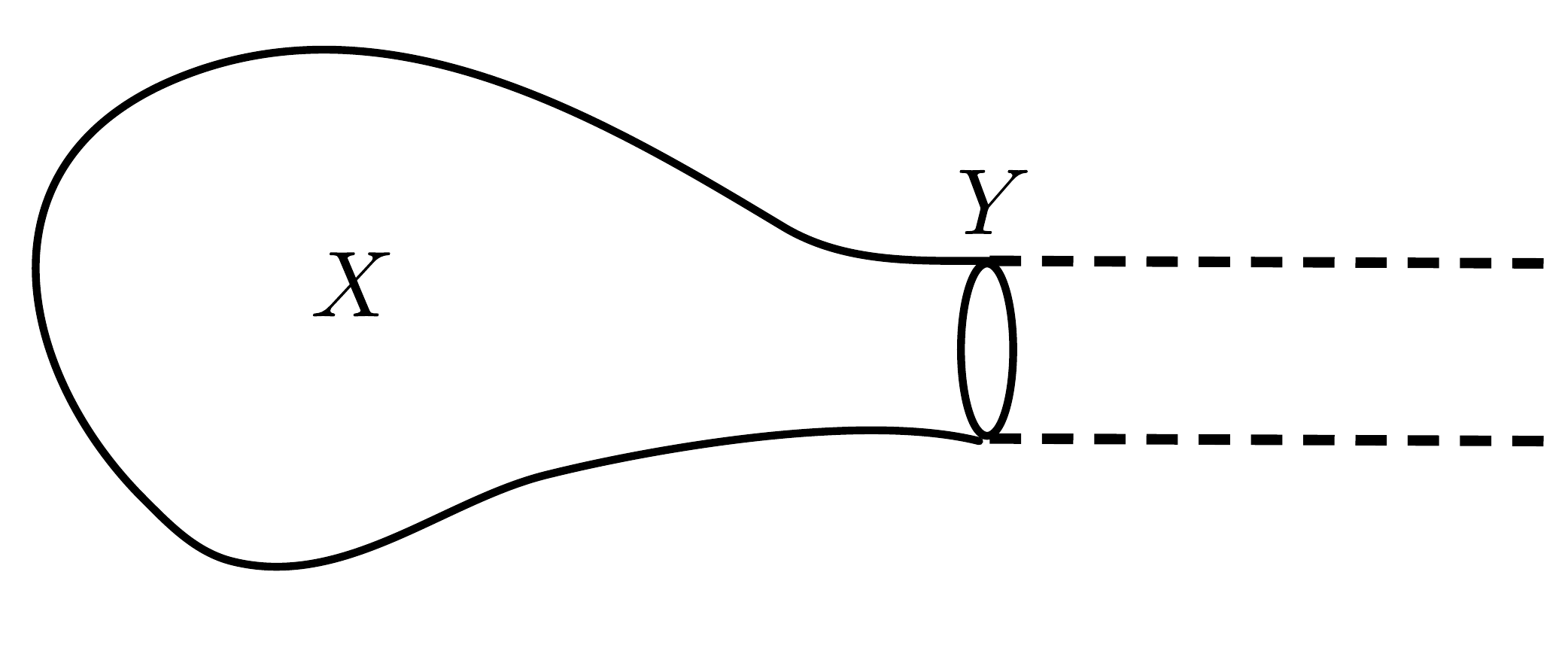}
\caption{Extension $\widehat X$ of the manifold $X$ by a semi-infinite cylindrical region which is represented by the dashed lines. \label{fig:long}}
\end{figure}

There is an intuitive way to understand why the APS boundary condition corresponds to the vacuum state. 
Suppose we attach a semi-infinitely long cylindrical region 
$[0,\infty) \times Y$ to $X$ as in Figure~\ref{fig:long}. 
We denote the extended manifold without boundary as $\widehat X$.
Then, we can consider the Dirac operator $\cD_X$
on the space of square-normalizable sections on $\widehat X$. The APS boundary condition is such that zero modes on $X$
can be extended to square-normalizable zero modes on $\widehat X$.\footnote{
In this discussion, we are implicitly assuming that $\cD_Y$ has no zero modes.}
So, let us consider the physics of massless fermions on $\widehat X$. 
The Wick-rotation is possible when we use the usual Feynman $\i \epsilon$ prescription in Lorentz signature.
The $\i \epsilon$ has been introduced to select the vacuum state $\ket{\Omega}$ in the infinite time region $|t| \to \infty$.
Thus, in the Euclidean path integral on the manifold $\widehat X$, we automatically get the vacuum $\ket{\Omega}$ at $\tau \to +\infty$.
This is the physical reason that the APS boundary condition is realized by the vacuum state.

\section{The derivation of the APS index theorem}
In this section, we give a derivation of the APS index theorem by using the axial anomaly.

\subsection{The basic setup}
\begin{figure}
    \centering
    \includegraphics[width=0.5\textwidth]{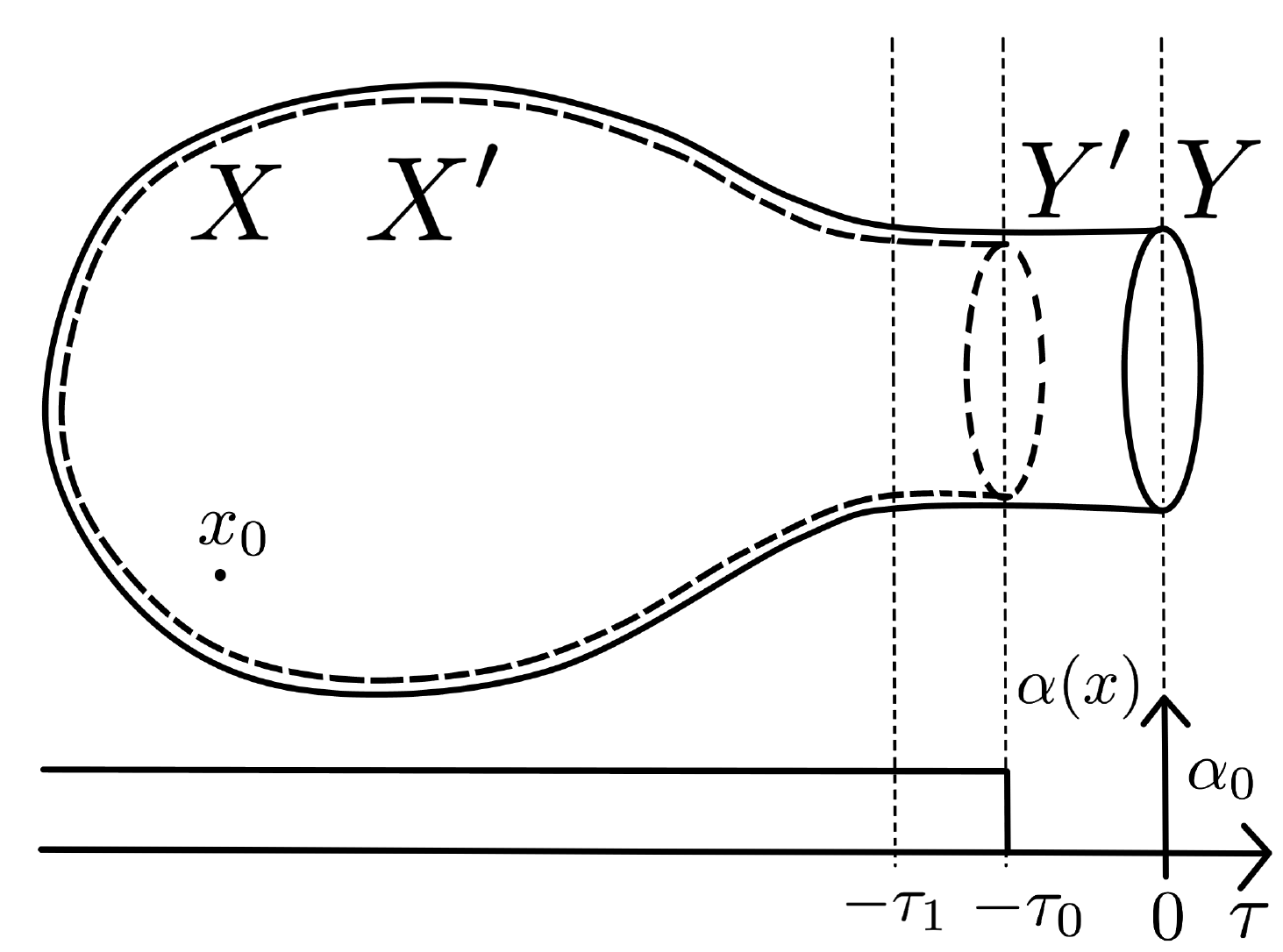}
    \caption{A manifold $X$ which has boundary $Y$. In the figure, $X'$ is the support of $\alpha(x) = \alpha_0\theta(-\tau-\tau_0)$, and its boundary
    is denoted as $Y'$.}
    \label{fig:ALL}
\end{figure}

\blue{
\paragraph{The Lagrangian}
}
We consider a manifold $X$ with boundary $\partial X = Y$. The situation is shown in Figure~\ref{fig:ALL}.
Near the boundary, $X$ is isometric to $(-\tau_1,0]\times Y$ for some $\tau_1>0$. The Dirac operator is
\beq
    \mathcal{D}_X := \i \slashed{D}_X = \i \gamma^\mu D_\mu.
\eeq
Near the boundary, we assume that the field strength 2-form of the gauge field $F$ and  
the curvature 2-form of the metric tensor $R$
do not contain $\d\tau$ and they are independent of $\tau$.

The Lagrangian in the Euclidean signature metric is
\beq
    \mathcal{L}_{\text{Euc}} = \overline{\Psi} \slashed{D}_X \Psi,
\eeq
which enters the Euclidean path integral as $e^{-S}$, where $S = \int \mathcal{L}_{\text{Euc}}$.
Notice the sign difference from $\mathcal{L}_{\text{Lor}}$.
This sign affects the APS index theorem.

\paragraph{Axial rotation.}
In the following, primes mean that we have performed axial rotation as
\beq
    \Psi'(x) =  e^{\i \alpha(x) \overline{\gamma} }\Psi(x) ,\quad 
	\overline{\Psi}'(x) = \overline{\Psi}(x) e^{\i \alpha(x) \overline{\gamma} } ,
\eeq
where $\alpha(x)$ is an axial transformation parameter which possibly has a nontrivial dependence on the position $x \in X$.
The Lagrangian is invariant under the global axial rotation $\alpha(x) = \alpha_0$.  
On the other hand, when $\alpha$ is not constant,
the Lagrangian changes by
\beq
    \mathcal{L}'_{\text{Euc}} = \mathcal{L}_{\text{Euc}} + \i j^\mu_A \partial_\mu \alpha(x),\label{eq:L=L+j}
\eeq
where $j^\mu_A$ is the axial current 
\beq
    j^\mu_A := \overline{\Psi}\gamma^\mu \overline{\gamma}\Psi.
\eeq
Its normalization is chosen so that its $\tau$ component, $j_A^\tau $, gives the charge \eqref{eq:Acharge} after 
setting $\overline\Psi = \Psi^\dagger \i \gamma^0 = \Psi^\dagger \gamma^\tau$ and integrating over $Y$.

\paragraph{Path Integral.} 
Formally, the partition function can be written as a product of eigenvalues of $\mathcal{D}_X$ as
\beq
    Z = \int [\mathcal{D}\Psi][\mathcal{D}\overline{\Psi}] e^{-S} = [\det \i \mathcal{D}_X] = \prod_a \i \lambda^{X}_a. \label{eq:Z}
\eeq
where $\lambda^{X}_a$ are eigenvalues of $\cD_X$ (which should not be confused with $\cD_Y$), 
and the product runs over all modes of $\cD_X$. 
If $\mathcal{D}_X$ has zero-modes, the partition function vanishes. 
Instead, we can calculate (unnormalized) expectation values of local operators $\mathcal{O}$ using the path integral:
\beq
    \vev{\mathcal{O}} = \int [\mathcal{D}\Psi][\mathcal{D}\overline{\Psi}] e^{-S}\mathcal{O}.
\eeq
The expectation value of $\cO$ does not vanish only if it has an appropriate charge under the axial $\U(1)$ symmetry
as we will discuss later.
 
Let us take a point $x_0 \in X$ such that it is far from the boundary. See Figure~\ref{fig:ALL}. 
 We also choose a local operator $\mathcal{O}(x_0)$ which has axial $\U(1)$ charge $2q$ and hence transforms as,
\beq
    \mathcal{O}'(x_0) = e^{2\i q\alpha(x_0)}\mathcal{O}(x_0). \label{eq:O=eqO}
\eeq
We will take $q$ such that the expectation value is nonzero, $\vev{\mathcal{O}(x_0)}\neq 0$.

\subsection{The axial rotation and the APS index theorem}

Now we can derive the APS index theorem by using the axial anomaly. The derivation
is very similar to that of the AS index theorem by Fujikawa, but we have to take into account the effect of the boundary.

\paragraph{Global axial rotation.}

First we consider global axial rotation 
\beq
\alpha(x) = \alpha_0,
\eeq
which is independent of $x$. 
As is well-known (see e.g. \cite[Sec.13.2]{Nakahara:2003nw})
the path integral measure changes as
\beq
    [\mathcal{D}\Psi'][\mathcal{D}\overline{\Psi}'] &= [\mathcal{D}\Psi][\mathcal{D}\overline{\Psi}] \exp\{-2\i \alpha_0 \index \mathcal{D}_X\}, \label{eq:D=DeInd}
\eeq
where $\index \mathcal{D}_X $ is the index of $\cD_X$ 
which is given in terms of the numbers of zero modes $n_\pm$ with positive and negative chirality $\overline{\gamma}\Psi =\pm\Psi$ as
$\index \mathcal{D}_X =n_+-n_- $.
This is well-known in the case of the usual AS index theorem, and the argument is unchanged in the current case of
the APS index theorem. We expand the field $\Psi$ into eigenmodes of $(\cD_X)^2$, and 
define the path integral measure as the product of integrals with respect to the coefficient of each mode.
The integral from the zero modes with positive and negative chirality contribute Jacobian factors $\exp(-\i n_+ \alpha_0)$ and 
$\exp(+\i n_- \alpha_0) $,
respectively. The extra minus sign in the exponent is due to the property of the Grassmannian integral
that the Jacobian is the inverse to that of the usual bosonic integral.

From (\ref{eq:O=eqO}) and (\ref{eq:D=DeInd}) we get
\beq
    \vev{\mathcal{O}(x_0)} 
        &= \int [\mathcal{D}\Psi][\mathcal{D}\overline{\Psi}] e^{-S}\mathcal{O}(x_0)\notag\\
		&= \int [\mathcal{D}\Psi'][\mathcal{D}\overline{\Psi}'] e^{-S'}\mathcal{O}'(x_0)\notag\\
		&=\int [\mathcal{D}\Psi][\mathcal{D}\overline{\Psi}]e^{-2\i \alpha_0\index \mathcal{D}_X} e^{-S}e^{2\i \alpha_0 q}\mathcal{O}(x_0)\notag\\
		&=\vev{\mathcal{O}(x_0)} \exp \left\{ -2\i \alpha_0[\index \mathcal{D}_X-q] \right\}, \label{eq:O=Oindq}
\eeq
where the second equality is just a variable change in the path integral. 
To have $\vev{\mathcal{O}(x_0)}\neq0$, we obtain a condition from (\ref{eq:O=Oindq}) that
\beq
    \index \mathcal{D}_X  =  q.	\label{eq:q=ind}
\eeq

\paragraph{Local axial rotation.}
Next we consider local axial rotation given by 
\beq
\alpha(x)=\alpha_0 \theta(-\tau-\tau_0)
\eeq
as shown in Figure~\ref{fig:ALL}. 
Notice in particular that $\alpha(x_0)=\alpha_0$. 
The reason that we take this configuration of $\alpha(x)$ is that it vanishes near the boundary.
Then the axial anomaly for this $\alpha(x)$ is insensitive to the presence of the boundary since the heat kernel method used by Fujikawa is local,
and hence we need not worry about the boundary. We can just apply the standard result of Fujikawa.
The path integral measure changes under the axial rotation by (see e.g. \cite[Sec.22.2]{Weinberg:1996kr})
\beq
    [\mathcal{D}\Psi'][\mathcal{D}\overline{\Psi}']
    &=[\mathcal{D}\Psi][\mathcal{D}\overline{\Psi}]\left[{\rm Det}\, (e^{\i \alpha\overline{\gamma}} ) \right]^{-2}\notag\\
    &=[\mathcal{D}\Psi][\mathcal{D}\overline{\Psi}]\exp\left[-2\i \Tr  \alpha\overline{\gamma}\right],
\eeq
where $\Tr$ means a trace over both the function space as well as spinor and gauge indices. 
The trace over the function space is formally calculated in the position basis:
\beq
    \Tr \alpha \overline{\gamma} 
        =\int \tr\overline{\gamma}\bra{x}\alpha\ket{x} 
        = \tr \overline{\gamma}\int_X\alpha(x) \delta(x-x),\label{eq:Trag}
\eeq
with $\tr$ denoting the trace over only spinor and gauge indices.

If $\alpha(x)$ were nonvanishing at the boundary, the above calculation would have been dangerous at the boundary. 
But we can use (\ref{eq:Trag}) because $\alpha$ vanishes near the boundary. Therefore we get
\beq
    [\mathcal{D}\Psi'][\mathcal{D}\overline{\Psi}']
    &=[\mathcal{D}\Psi][\mathcal{D}\overline{\Psi}] \exp\left\{-2\i \int_X \alpha \trg\right\}, \label{eq:D=DeG}
\eeq
where $G$ is formally given by
\beq
    \trg(x) = \tr{\overline{\gamma}}\delta(x-x).
\eeq
After regularization, it is given in terms of the background metric and gauge field. 

An explicit expression for $G(x)$ in terms of the curvature 2-forms $F$ and $R$ of the gauge field and the metric tensor is well-known.
For example, if we consider an even-dimensional manifold $d=2n$ and take $\Ch = \i^{-n} \gamma^1 \gamma^2 \cdots \gamma^{2n}$, we have
(see e.g.\cite[Sec.12.10,13.2]{Nakahara:2003nw})
\beq
    \trg & = \left. \hat{A}(R) \ch(F) \right|_{2n}, \label{eq:G=Ach}
\eeq
where
\beq
    \hat{A}(R) := \sqrt{\det \frac{\i R/(4\pi)}{\sinh \i R/(4\pi)}}, \qquad
    \ch(F) := \tr \exp \left(\frac{\i F}{2\pi}\right).
\eeq
However, we need not assume that the manifold $X$ is even-dimensional. 
The APS index theorem is valid even on odd-dimensional manifolds and/or nonorientable manifolds.
Such cases are important for practical applications (see e.g. \cite{Witten:2015aba,Tachikawa:2018njr}).
What we need is (i) a first order elliptic partial differential operator $\cD_X$ which is self-adjoint, and (ii) a $\bZ_2$-grading $\Ch$ which anticommutes with $\cD_X$.
Thus we proceed abstractly without assuming the explicit formula for $G$.

We have
\beq
    \int_X \alpha \trg 
    = \alpha_0\int_{X'}  \trg 
    = \alpha_0\int_{X} \trg ,
\eeq
where $X'$ is the support of $\alpha(x)$. (see Figure~\ref{fig:ALL}.)
The second equality is due to the fact that 
we have assumed that $F$ and $R$ do not contain $\d\tau$ in the region $-\tau_0 \leq \tau \leq 0$ and hence 
$G$ vanishes in this region.

Using (\ref{eq:L=L+j}), (\ref{eq:O=eqO}) and (\ref{eq:D=DeG}) we get
\beq
    \vev{\mathcal{O}(x_0)} 
		&= \int [\mathcal{D}\Psi'][\mathcal{D}\overline{\Psi}'] e^{-S'}\mathcal{O}'(x_0)\notag\\
		&= \int [\mathcal{D}\Psi][\mathcal{D}\overline{\Psi}]e^{-2\i \int_X\alpha \trg } e^{-S- \i\int_X j^\mu_A\partial_\mu\alpha} e^{2\i\alpha(x_0) q}\mathcal{O}(x_0)\notag\\
		&=\int [\mathcal{D}\Psi][\mathcal{D}\overline{\Psi}] e^{-S}\mathcal{O}(x_0)
		\exp\left\{-2\i \left[\alpha_0\int_{X} G+\frac{1}{2}\int_X j^\mu_A\partial_\mu\alpha-\alpha_0 q\right]\right\}.
\eeq
Since $\alpha(x) = \alpha_0 \theta(-\tau-\tau_0)$, we get
\beq
    j^\mu_A \partial_\mu \alpha(x) = j^\tau_A\partial_\tau \alpha(x) = -j^\tau_A\delta(\tau+\tau_0)\alpha_0.
\eeq
We also have
\beq
    \int_{Y'}j_A^\tau &= \int_{Y}j_A^\tau = Q_A,
\eeq
where $Y'$ is the boundary of $X'$ (See Figure~\ref{fig:ALL}).
The reason for the equality between $ \int_{Y'}j_A^\tau $ and $ \int_{Y}j_A^\tau $ is as follows. The divergence of the current $\nabla_\mu j^\mu_A$
is proportional to the axial anomaly $G$. However, $G$ vanishes in the region $-\tau_0 \leq \tau \leq 0$. Therefore, $j^\mu_A$ is conserved in this region
and the equality is just the conservation of the charge. 

From the above result, we obtain
\beq
    \vev{\mathcal{O}(x_0)} 
        &=\int [\mathcal{D}\Psi][\mathcal{D}\overline{\Psi}] e^{-S}
		\exp\left\{-2\i \alpha_0\left[\int_X G-\frac{1}{2} Q_A- q\right]\right\}\mathcal{O}(x_0). \label{eq:O=OeQ}
\eeq

\paragraph{The APS theorem.}
Now we can apply the result of Sec.~\ref{sec:eta}. The axial charge $Q_A$ in the above equation is defined on the boundary and hence it acts on the boundary state.
We have discussed that the APS boundary condition corresponds to the vacuum state $\bra{\Omega}$. 
The axial charge of the vacuum is computed in Sec.~\ref{sec:eta} and it is given by
\beq
\bra{\Omega}Q_A = \bra{\Omega}(-2\eta(\cD_Y) ).
\eeq
Substituting this result into \eqref{eq:O=OeQ}, we obtain 
\beq
    \vev{\mathcal{O}(x_0)} = \vev{\mathcal{O}(x_0)} \exp\left\{-2\i \alpha_0\left[\int_{X} G +\eta(\cD_Y) - q\right]\right\}.
\eeq
Since $\vev{\mathcal{O}(x_0)}\neq0$, we need to have
\beq
q= \int_X G + \eta(\cD_Y). \label{eq:q=G+eta}
\eeq

Combining \eqref{eq:q=ind} and \eqref{eq:q=G+eta}, we get 
\beq
 \index \mathcal{D}_X =  \int_X G + \eta(\cD_Y).
\eeq
This is the APS index theorem.

\subsection{Generalization to other boundary conditions} \label{sec:general}
So far we have discussed the APS index theorem when we impose the APS boundary condition. However,
our argument can be easily generalized to other boundary conditions as far as $\cD_X$ is self-adjoint under that boundary condition. 
We just replace the vacuum $\bra{\Omega}$ in the final state to another state $\bra{\beta}$ with a certain property
which we discuss below. 
We also denote the boundary condition as $\beta$. 
The relation between physical states and boundary conditions are mentioned in \eqref{eq:BCandState}.

The index depends on the boundary condition, so let us denote it as $\index(\cD_X, \beta)$.
Also, we denote the axial $\U(1)$ charge of the state $\ket{\beta}$ as $-2\eta(\cD_Y, \beta)$.
Then, a generalized APS index theorem is
\beq
 \index(\cD_X, \beta) =  \int_X G + \eta(\cD_Y, \beta).
\eeq
There is no change in the bulk contribution $G$. 

We are interested in the index theorem, so the boundary condition $\beta$ must be chosen in such a way
that the Dirac operator $\cD_X$ with the boundary condition is self-adjoint.
The self-adjointness is not a physical requirement, but it is necessary only because we are interested in the index theorem.
This requires, among other things, the following. When we try to show that the operator $\cD_X$ is self-adjoint,
we need to use integration by parts. The surface term which appears in this integration by parts must vanish
for $\cD_X$ to be self-adjoint. Thus we need to choose boundary conditions such that the surface term vanishes.

We define an inner product between two sections $\Psi_1$ and $\Psi_2$ as $(\Psi_1, \Psi_2) = \int_X \o \Psi_1 \Psi_2$,
where the bar on $\o \Psi_1$ is the adjoint of the section $\Psi_1$ in the Euclidean space, which should not be confused with
the hermitian conjugate on operators acting on the Hilbert space.
The surface term in the integration by parts $(\Psi_1, \cD_X \Psi_2) = (\cD_X\Psi_1,  \Psi_2) $ is $\int_Y \i \o \Psi_1 \gamma^\tau \Psi_2$.

The vanishing of the surface term is achieved by what is called generalized APS boundary conditions. 
See \cite{Yonekura:2016wuc} for the physical state vectors realizing it.

Here we just state how we can realize examples of generalized APS boundary conditions. 
Recall the expansion \eqref{eq:PsiExpand}, and define operators $S_i := A_{+,i} A_{-,i}^\dagger$.
When $\lambda_i>0$, the vacuum $\ket{\Omega}$ is annihilated by both $ A_{+,i}$ and $ A_{-,i}^\dagger$.
Then we can create the excited modes for both $ A_{+,i}^\dagger$ and $ A_{-,i}$. simultaneously by acting $S_i^\dagger$.
On the other hand, when $\lambda_i < 0$, the vacuum $\ket{\Omega}$ is annihilated by both $ A_{+,i}^\dagger$ and $ A_{-,i}$.
Then we can create the excited modes for both $ A_{+,i}$ and $ A_{-,i}^\dagger$ simultaneously by acting $S_i$.
States created by acting $S_i$ or $S_i^\dagger$ to the vacuum are generalized APS boundary conditions.
As a consistency check, notice that $S_i $ and $S_i^\dagger$ have axial charge $\pm 2$, so by acting these operators the value of $Q_A$ changes
by even numbers. Then $\eta(\cD_Y, \beta)$, which is defined as the value of $-\frac12 Q_A$ on the state $\bra{\beta}$, 
changes by integers. It is consistent with the fact that $ \index(\cD_X, \beta) $ is an integer. 

Finally let us comment on zero modes of $\cD_Y$. So far we have assumed that $\lambda_i \neq 0$.
Now let us consider zero modes $\lambda_\alpha=0$, and denote their coefficients as $A_{+,\alpha}$ and $A_{-,\alpha}$.
Then, the states must be annihilated by both $(A_{+,\alpha}, A_{-,\alpha}^\dagger)$, or annihilated by both $(A_{+,\alpha}^\dagger, A_{-,\alpha})$.
These are the APS boundary conditions when $\cD_Y$ has zero modes.

\section*{Acknowledgements}
The work of KY is supported in part by JSPS KAKENHI Grant-in-Aid (Wakate-B), No.17K14265.

\bibliographystyle{ytphys}
%\baselineskip=.95\baselineskip
\bibliography{ref}

\end{document}